\documentclass[graybox]{svmult}

\usepackage{mathptmx}       
\usepackage{helvet}         
\usepackage{courier}        
\usepackage{makeidx}         
\usepackage{graphicx}        
\usepackage{multicol}        
\usepackage[bottom]{footmisc}

\makeindex             

\begin{document}

\title*{One-Dimensional Flow Spectra and Cumulative Energy from Two Pipe Facilities}
\author{El-Sayed Zanoun, Emir \"Ong\"uner, Christoph Egbers, Gabriele Bellani, and Alessandro Talamelli}

\institute{El-Sayed Zanoun \at Department of Aerodynamics and Fluid Mechanics, Brandenburg University of Technology, 03046-Cottbus, Germany, \email{el-sayed.zanoun@b-tu.de}
\and Emir \"Ong\"uner \at Institute of Aerodynamics and Flow Technology, German Aerospace Center (DLR), G\"{o}ttingen, Germany, \email{emir.oenguener@b-tu.de}
\and Christoph Egbers \at Department of Aerodynamics and Fluid Mechanics, Brandenburg University of Technology, Germany, \email{christoph.egbers@b-tu.de}  
\and Gabriele Bellani \at Center for Industrial Research in Aerospace Engineering, University of Bologna, 47100-Forli, Italy, \email{gabriele.bellani2@unibo.it}
\and Alessandro Talamelli \at Center for Industrial Research in Aerospace Engineering, University of Bologna, 47100-Forli, Italy, \email{alessandro.talamelli@unibo.it}
}

\maketitle

\abstract*{Experiments have been conducted to assess the sizes and energy fractions of structure in fully developed turbulent pipe flow regime in two pipe facilities, ColaPipe at BTU Cottbus-Senftenberg, and CICLoPE at University of Bologna, for shear Reynolds number in the range $2.5\cdot{10^3}\le{\mathrm{Re_{\tau}}}\le{{3.7\cdot{10^4}}}$, utilizing a single hot-wire probe. Considerations are given to the spectra of the streamwise velocity fluctuations, and to large scale motions and their energy contents from the pipe near-wall to centerline. The analysis of the velocity fluctuations revealed a Reynolds-number dependent inner peak at a fixed wall normal location, however, an outer peak seems not to appear that might be attributed either to low Reynolds number effect or not high enough spatial resolution of hot-wire probe, motivating further study utilizing  nanoscale probes. Sizes of the large scale, and very large scale structures were estimated to have wavelengths of 3$R$, and 20$R$ at high Reynolds number, srespectively. The fractional energy contents in wavelengths associated with the large scale motions at various wall normal locations showed maximum contribution to the turbulent kinetic energy near the outer limit of the logarithmic layer.} 

\abstract{Experiments have been conducted to assess the sizes and energy fractions of structure in fully developed turbulent pipe flow regime in two pipe facilities, ColaPipe at BTU Cottbus-Senftenberg, and CICLoPE at University of Bologna, for shear Reynolds number in the range $2.5\cdot{10^3}\le{\mathrm{Re_{\tau}}}\le{{3.7\cdot{10^4}}}$, utilizing a single hot-wire probe. Considerations are given to the spectra of the streamwise velocity fluctuations, and to large scale motions and their energy contents from the pipe near-wall to centerline. The analysis of the velocity fluctuations revealed a Reynolds-number dependent inner peak at a fixed wall normal location, however, an outer peak seems not to appear that might be attributed either to low Reynolds number effect or not high enough spatial resolution of hot-wire probe, motivating further study utilizing  nanoscale probes. Sizes of the large scale, and very large scale structures were estimated to have wavelengths of 3$R$, and 20$R$ at high Reynolds number, srespectively. The fractional energy contents in wavelengths associated with the large scale motions at various wall normal locations showed maximum contribution to the turbulent kinetic energy near the outer limit of the logarithmic layer.} 

\section{Introduction}
\label{sec:1}
Turbulent large scale structures in pipe facilities at high Reynolds numbers are of practical importance in terms of their fractional contributions to Reynolds stresses and energy budgets. For decades, understanding such turbulent flow structures have been of interest by, e.g., Townsend (1976), Kim, Adrian (1999), Marusic et al. (2010), Morrison et al. (2016), and Jim\'enez (2018). Nevertheless, conceret definition of origin, natures, evolutions, and sizes of such large scale structures are still under debat, in particular, at high Reynolds numbers. This short contribution aims, therefore, at characterizing experimentally the following few features of such large scale structure in two pipe facilities at high Reynolds numbers:
\begin{itemize}
\item {Scaling the streamwise Reynolds stress and examining its inner and outer peaks.}
\item {Scaling spectra of the streamwise velocity fluctuations ($\Phi_{uu}$).}
\item {Estimating sizes of the large and very large scale motions, i.e. LSM $\&$ VLSM.}
\item {Examining contribution of the large scale motions to turbulent kinetic energy.}
\end{itemize}

\section{Facilities and Measuring Techniques}
\label{sec:2}

Current measurements in CoLaPipe $\&$ CICLoPE aimed at investigating the streamwise energy spectra over a wide range of the shear Reynolds number ($Re_{\tau}=u_{\tau}R/\nu$), where  $u_{\tau}$ is the wall friction velocity, $R$ is the pipe radius, and $\nu$ is the kinematic viscosity. The CoLaPipe is closed return facility, located at Brandenburg University of Technology, Germany, to carry out measurements for $1.5\times{10^3} \leq Re_{\tau} \leq 19\times{10^3}$. The facility provides air with $\approx$78m/s maximum velocity, having turbulence level less than $0.5\%$, see K\"onig (2014). It has two pipe sections, both made out of high-precision smooth acrylic glass, having inner pipe diameters of $190\pm{0.23}$mm and $342\pm{0.35}$mm, with ${136}$, and ${77}$ length-to-diameter-ratio ($L/D$) for the suction, and the return sections, respectively. The CICLoPE is also closed return facility, located at University of Bologna, Italy, for a range of ${10^3} \leq Re_{\tau} \leq 4\times{10^4}$. The CICLoPE facility provides air with 60 m/s maximum velocity, see Fiorinin (2017). The CICLoPE has a pipe section of $900\pm{0.2}$mm as inner diameter, and 111.5m total length, i.e. $L/D={124}$. It is to  note that both facilities are equipped with water coolers.

Measurements in CICLoPE facility have been carried out using Dantec Streamline 90N10 CTA, and Dantec 55P11 commercial probes in addition to custom-made Platinum single-wire probes, see Fiorinin (2017). On the other hand, in CoLaPipe facility, all measurements have been conducted using Dantec Multichannel-CTA 54N81 with commercial Dantec boundary layer probe, Model 55P53. The sampling frequencies were set to 60 kHz $\&$ 20 kHz with a low-pass filter at $f_{LP}$ =30 kHz $\&$ 10 kHz, for the CICLOPE and CoLaPIpe, respectively, and samples were acquired over 120 s at every measuring point. It is worth noting that the mean pressure gradient along both pipe test sections was used to estimate the wall friction velocity ($u_{\tau}$).

\section{Results and Discussions}
\label{sec:3}

Figure 1 illustrates the inner scaling of the streamwise mean, and fluctuating velocity profiles from experiments and simulations. The local scaling for velocity, and wall normal distance were carried out using the wall friction velocity ($u_{\tau}$), and the viscous length scale ($\ell_c=\nu/u_{\tau}$), respectively. The mean velocity profiles in Fig. 1(left) show good collapse, and a satisfactory agreement with the logarithmic line,  $U^+=1/\kappa{\mathrm{ln}{(y^+)}}+B$, where $\kappa=0.384$ and $B=4.43$ proposed by Zanoun et al. (2007). The figure also presents samples for the experimental streamwise velocity fluctuations at $Re_\tau=2675$ $\&$ 11000. Plausible agreement between  experiments for $Re_\tau=2675$ with DNS data from Ahn et al. (2015) is observable. For high enough Reynolds number, back to 1976, Townsend showed that the streamwise turbulence intensity behave logarithmically, $u'^{+2}=B_2-A_2{\mathrm{ln}{(y^+/Re_{\tau})}}$, in the interior part of the inertial region. A clear logarithmic behavior for $u'^{+2}$ is being seen in Fig. 1(left) for $Re_\tau=11000$ with $A_2=1.25$, and $B_2=1.61$.
 
\begin{figure}[h]
\sidecaption
\centering
\includegraphics[scale=0.26]{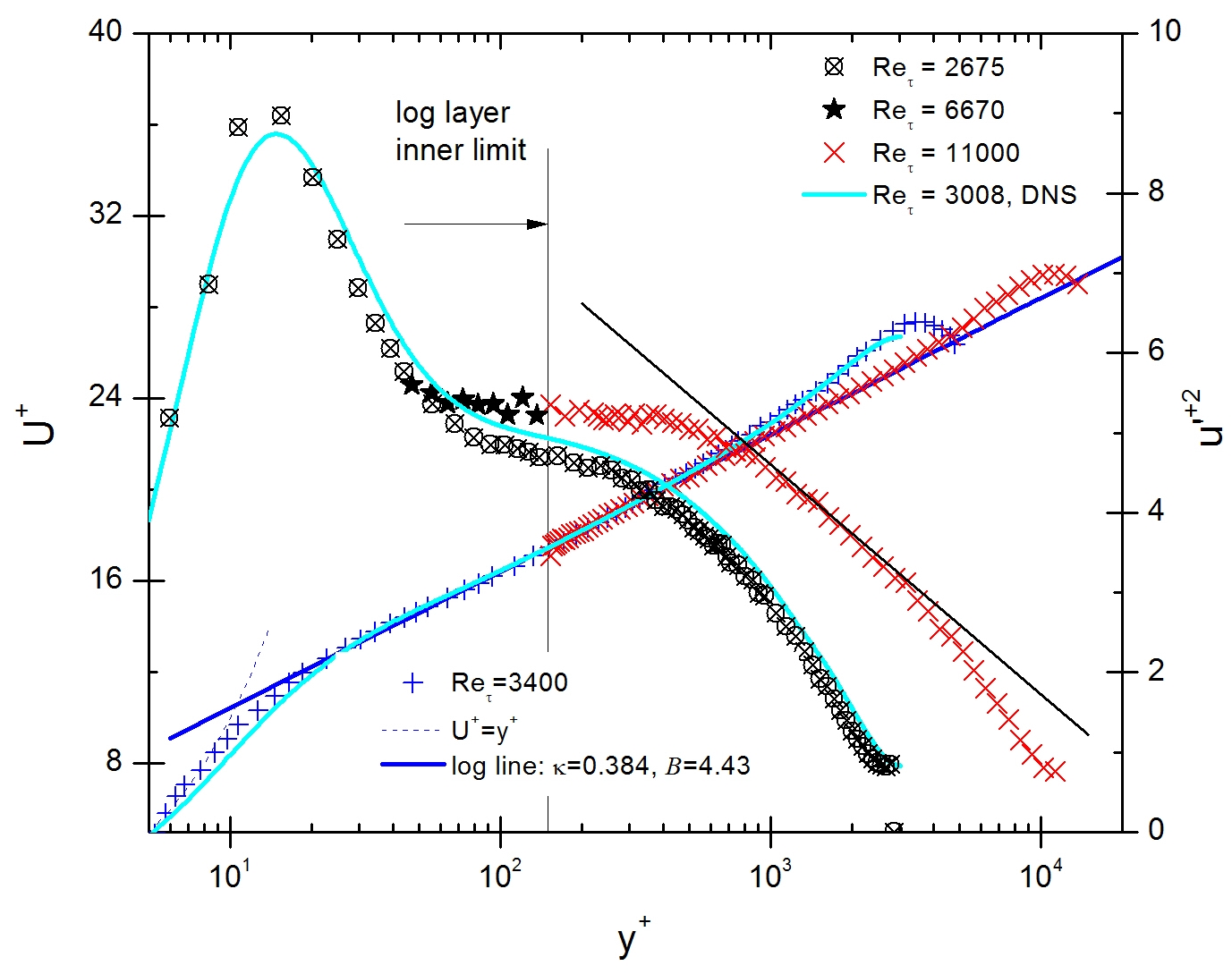}
\includegraphics[scale=0.27]{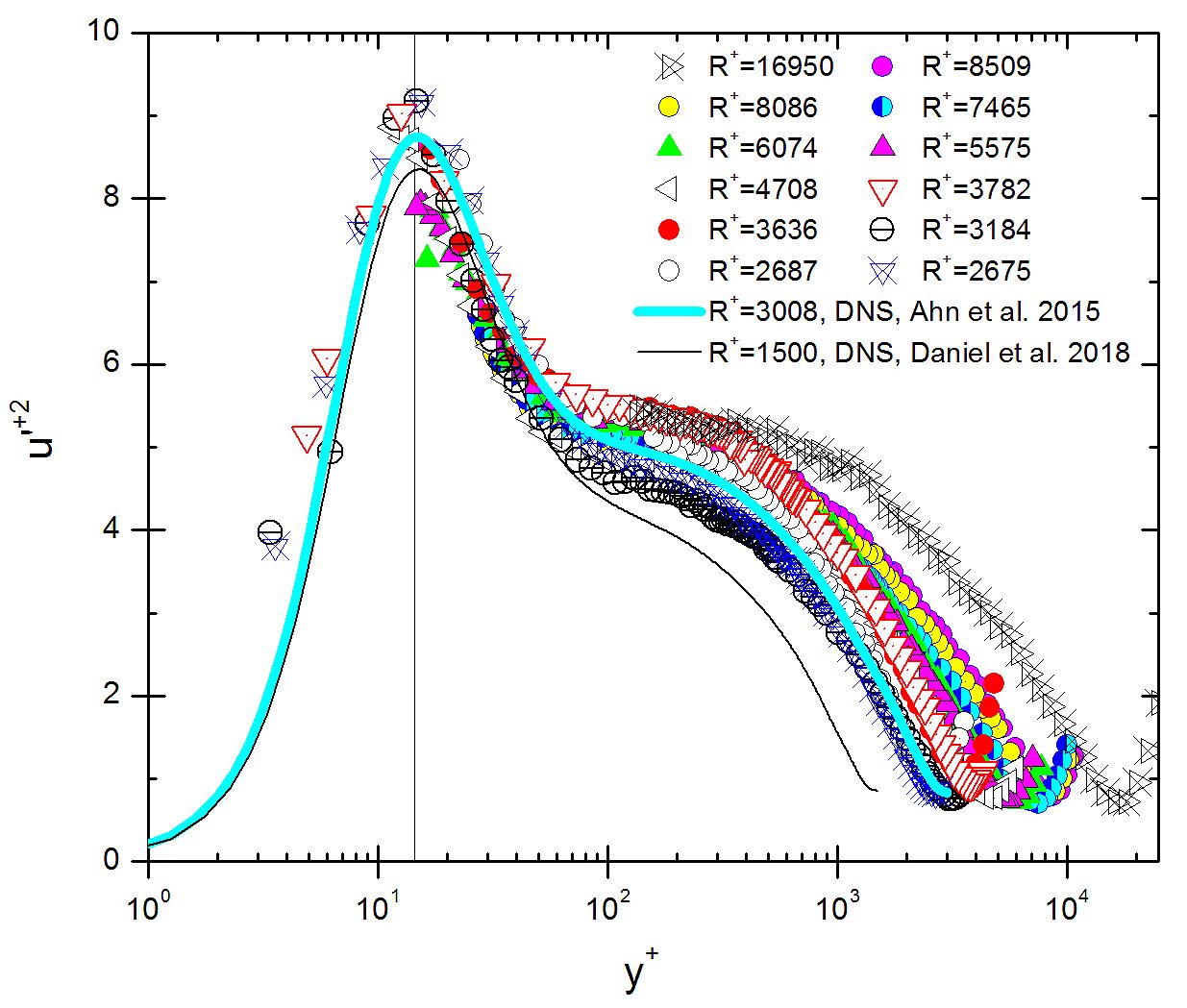}
\caption{Inner scaling of the streamwise mean, and fluctuating velocity profiles from experiments (CoLaPipe) and simulations (DNS by Refs. 9 $\&$ 11).}
\label{4800_37000_both_pmps}    
\end{figure}

Figure 1(right) illustrates the inner scaling of all $u'^{+2}$ from experiments, and two available DNS data sets [Ahn et al. (2015), Daniel et al. (2018)] versus the wall normal location $y^+$. Focusing on the region close to the wall, the data do collapse and follow the classical scaling with Re-dependent inner peak, located at $y^+\approx{15}$ in good agreement with predictions utilizing Hutchins et al. (2009) empirical formula. The location of the inner peak observed is also in correspondence with the location where the turbulent kinetic energy production reaches its maximum, Zanoun and Durst (2009). On the other hand, an outer peak is hardly observable in Fig. 1(right) which is attributed either to low Reynolds number effects or not high enough spatial resolution of hot wire probe, however, a plateau is being clear along the overlap region. Further research is, therefore, being motivated utilizing  NSTAPs probes to see whether the second extremum a real peak or does it form a shoulder or plateau.

Selected samples from pre-multiplied spectra (${k_x}{\Phi_{uu}/{{u_\tau}^2}}$) of the streamwise velocity fluctuations presented in Fig. 2 for $Re_\tau=3500$ (a,b) $\&$ $Re_\tau=4800$ (c,d) highlight questions raised earlier in section 1, utilizing CoLaPipe and CICLoPE, respectively. In outer scaling, Figs. 2(a) $\&$ (c) show satisfactory similarity at low wavenumbers, while in wall-normal scaling, data in Figs. 2(b) $\&$ (d) show similarity for moderately high wavenumbers. One might speculate that such spectra represents a footprint for the large and very large scale motions (LSM $\&$ VLSM ) observed in both facilities with two discernible length sales, one at low, and the other at moderate wavenumbers associated with the VLSM, and the LSM, respectively. The dashed line presented in both Figs. 2(b) $\&$ (d) represents a proposed logarithmic correction by del \'Alamo et al. (2004) to the $k_x^{-1}$ spectrum given by ${k_x}{\Phi_{uu}}=\beta {u_{\tau}^2} \mathrm{log}(2\pi\alpha^2/(k_xy))$, with $\alpha\approx{2}$ and $\beta \approx{0.2}$, indicating good agreement with experiments within the range $0.63<k_xy< 6.3$. 

\begin{figure}[h]
\sidecaption
\centering
\includegraphics[scale=0.35]{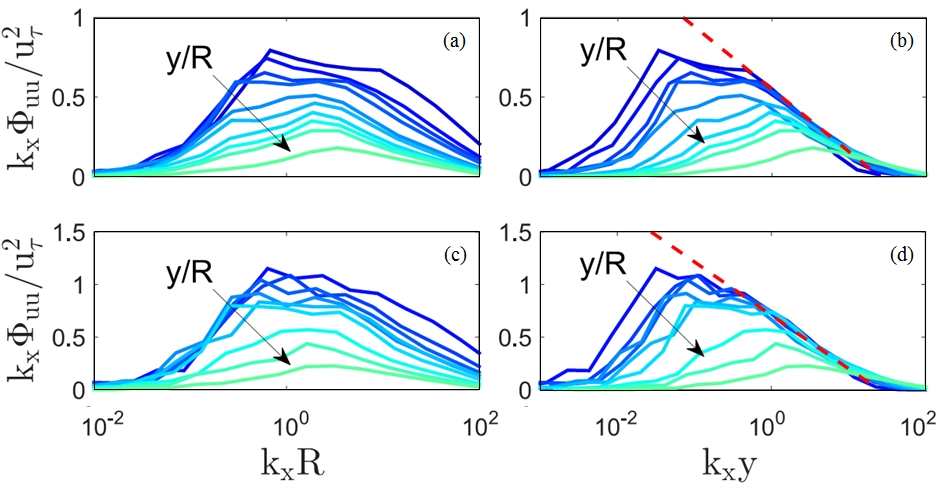}
\caption{Outer (Left), and wall normal (right) scaling of pre-multiplied spectra versus normalized wall distance for y/R= 0.05, 0.1, 0.15, 0.2, 0.3, 0.4, 0.5, 0.6, 0.7, 1.0 (arrows indicate decreasing trend), CoLaPIpe: (a,b) $Re_\tau=3500$  $\&$ CICLOPE: (c,d) $Re_\tau=4800$.}
\label{4800_37000_both_pmps}    
\end{figure}

Based on spectral peaks observed in Fig. 2, sizes of the large-scale structures were estimated and presented in Fig. 3. Very large scale motions (VLSM) start within the buffer layer and grow through the inertial sublayer, reaching maximum wavelength of $\lambda_x\approx{19}R$ for $Re_\tau\approx17000$, and $\lambda_x\approx{20}R$ for $Re_\tau\ge{20000}$ at half of the pipe radius, i.e. outside the logarithmic layer, see Fig. 3(a). A sudden drop in the normalized wavelength of the very large scale motion is observed at $y/R\approx{0.5}$ in Fig. 3(a) due to merging of the VLSM and the LSM forming one flow structure beyond $y/R=0.5$. Such structure (LSM) presented in Fig. 3(b) spans the pipe cross-section from a location close to the pipe wall to the centerline of the pipe with a wavelength of $\lambda_x\approx{3R}$ in good agreement with Kim, and Adrian (1999). Figure 3 represents quantitative comparison and good summary for sizes of the large scale motions between the two pipe facilities over a wide range of Reynolds number.

\begin{figure}[h]
\sidecaption
\centering
\includegraphics[scale=.34]{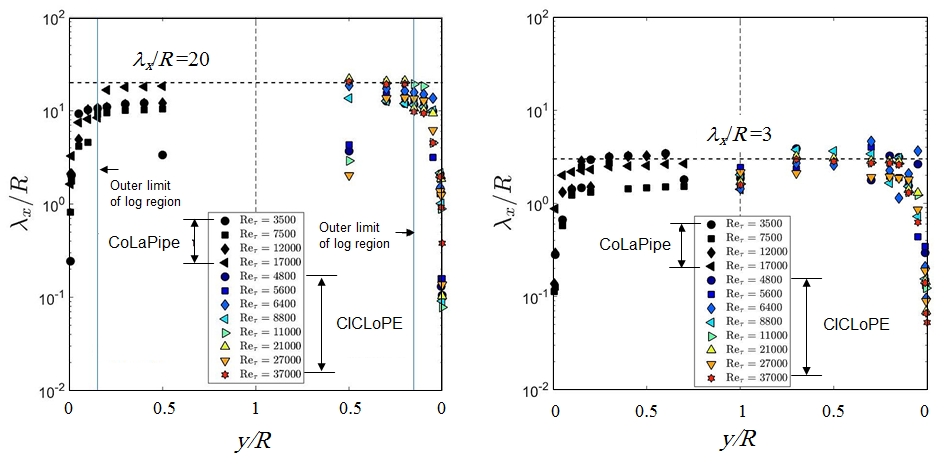}
\caption{Dimensionless wavelengths of the VLSM (left) $\&$ LSM (right), for a wide range of Reynolds number in both facilities. In each subplot, CoLaPipe is the left half, and CICLoPE is the right half.}
\label{VLSM_inner}    
\end{figure}

The cumulative energy for wall normal locations of $0.005\le{y/R}\le{1}$ at $\mathrm{Re_{\tau}}={3500-37000}$ is presented in Fig. 4, addressing the contribution of both the LSM and the VLSM to the turbulent kinetic energy. The contribution of such structures within a range of wavelengths to energetic the flow is examined via the distribution of the streamwise cumulative energy [$\Upsilon_{uu}=1-\sum_{0}^{k}\Phi_{uu}(k)/\sum_{0}^{k_{max}}\Phi_{uu}(k)$] as a function of the normalized streamwise wavelength $\lambda_x/R$. For instance, one would estimate from data presented in Fig. 4 that at a wall normal location $y/R=0.2$, flow structures with wavelengths greater than 3, 10, and 20 pipe radii have fractional contributions around $57\%$, $35\%$, and $22\%$, respectively, to the turbulent kinetic energy. This location is in correspondence with the top of the logarithmic layer. 

\begin{figure}[h]
\sidecaption
\centering
\includegraphics[scale=.36]{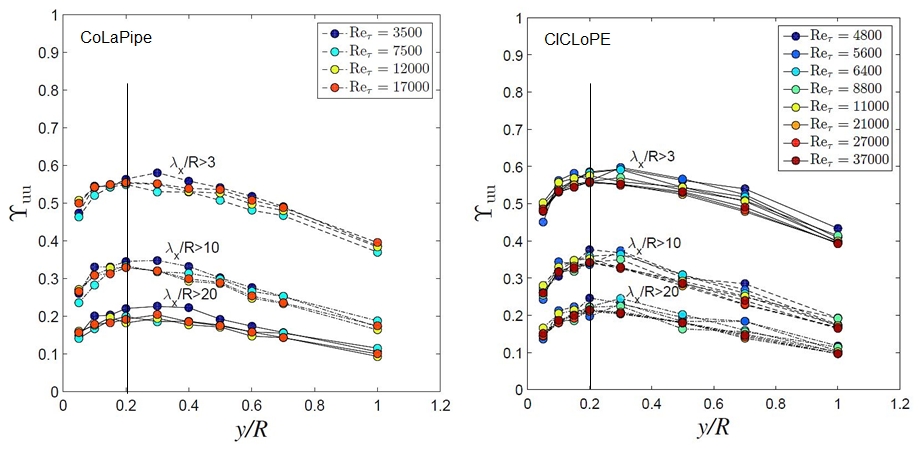}
\caption{Fractional contributions of the large scale motions to the kinetic energy in both facilities, (left) CoLaPipe, and (right) CICLoPE.}
\label{VLSM_outer}    
\end{figure}

\section{Conclusions and Future Work}
\label{sec:4}

An experimental study of the large scale motions have been carried out in two new experimental pipe facilities where flow was assured to be fully developed for $\mathrm{Re_{\tau}}={2500-37000}$. Results using a single hot-wire probe showed very large and large scale motions with wavelengths $\approx$20$R$, and $\approx{3R}$, for $Re_{\tau}\ge{10^4}$ at half of the pipe radius, and spans from the wall to the centerline of the pipe, respectively. Around $57\%$, $35\%$ and $22\%$ of the kinetic energy are attributed to length scales greater than 3, 10, and $\&$ 20 pipe radii, respectively, around the top of the inertial sublayer. Further research is, however, being motivated utilizing  NSTAPs probes.\\


{\bf Acknowledgements} This research is funded via the German Research Foundation (DFG) as part of the FOR1182 and SPP1881 project. Support received from the European High performance Infrastructures in Turbulence (EUHIT) is appreciated.

%
%
%

\end{document}